\documentstyle[psfig,epsfig]{mn}

\newif\ifAMStwofonts
\ifoldfss
  \ifCUPmtlplainloaded \else
    \NewTextAlphabet{textbfit} {cmbxti10} {}
    \NewTextAlphabet{textbfss} {cmssbx10} {}
    \NewMathAlphabet{mathbfit} {cmbxti10} {} % for math mode
    \NewMathAlphabet{mathbfss} {cmssbx10} {} %  "   "    "
  \fi
  \ifAMStwofonts
    \ifCUPmtlplainloaded \else
      \NewSymbolFont{upmath} {eurm10}
      \NewSymbolFont{AMSa} {msam10}
      \NewMathSymbol{\upi}     {0}{upmath}{19}
      \NewMathSymbol{\umu}     {0}{upmath}{16}
      \NewMathSymbol{\upartial}{0}{upmath}{40}
      \NewMathSymbol{\leqslant}{3}{AMSa}{36}
      \NewMathSymbol{\geqslant}{3}{AMSa}{3E}

       \let\le=\leqslant
       
    \fi
  \fi
\fi % End of OFSS

\ifnfssone
  \newmathalphabet{\mathit}
  \addtoversion{normal}{\mathit}{cmr}{m}{it}
  \addtoversion{bold}{\mathit}{cmr}{bx}{it}
  \newmathalphabet{\mathbfit} % math mode version of \textbfit{..}
  \addtoversion{normal}{\mathbfit}{cmr}{bx}{it}
  \addtoversion{bold}{\mathbfit}{cmr}{bx}{it}
  \newmathalphabet{\mathbfss} % math mode version of \textbfss{..}
  \addtoversion{normal}{\mathbfss}{cmss}{bx}{n}
  \addtoversion{bold}{\mathbfss}{cmss}{bx}{n}
  \ifAMStwofonts
    \ifCUPmtlplainloaded \else
      \UseAMStwoboldmath
      \makeatletter
      \new@mathgroup\upmath@group
      \define@mathgroup\mv@normal\upmath@group{eur}{m}{n}
      \define@mathgroup\mv@bold\upmath@group{eur}{b}{n}
      \edef\UPM{\hexnumber\upmath@group}
      \new@mathgroup\amsa@group
      \define@mathgroup\mv@normal\amsa@group{msa}{m}{n}
      \define@mathgroup\mv@bold\amsa@group{msa}{m}{n}
      \edef\AMSa{\hexnumber\amsa@group}
      \makeatother
      \mathchardef\upi="0\UPM19
      \mathchardef\umu="0\UPM16
      \mathchardef\upartial="0\UPM40
      \mathchardef\leqslant="3\AMSa36
      \mathchardef\geqslant="3\AMSa3E

       \let\le=\leqslant

    \fi
  \fi
\fi % End of NFSS release 1

\ifnfsstwo
  \DeclareMathAlphabet{\mathbfit}{OT1}{cmr}{bx}{it}
  \SetMathAlphabet\mathbfit{bold}{OT1}{cmr}{bx}{it}
  \DeclareMathAlphabet{\mathbfss}{OT1}{cmss}{bx}{n}
  \SetMathAlphabet\mathbfss{bold}{OT1}{cmss}{bx}{n}
  \ifAMStwofonts
    \ifCUPmtlplainloaded \else
      \DeclareSymbolFont{UPM}{U}{eur}{m}{n}
      \SetSymbolFont{UPM}{bold}{U}{eur}{b}{n}
      \DeclareSymbolFont{AMSa}{U}{msa}{m}{n}
      \DeclareMathSymbol{\upi}{0}{UPM}{"19}
      \DeclareMathSymbol{\umu}{0}{UPM}{"16}
      \DeclareMathSymbol{\upartial}{0}{UPM}{"40}
      \DeclareMathSymbol{\leqslant}{3}{AMSa}{"36}
      \DeclareMathSymbol{\geqslant}{3}{AMSa}{"3E}

       \let\le=\leqslant

    \fi
  \fi
\fi % End of NFSS release 2

\ifCUPmtlplainloaded \else
  \ifAMStwofonts \else % If no AMS fonts
    \def\upi{\pi}
    \def\umu{\mu}
    \def\upartial{\partial}
  \fi
\fi
\newcommand{\be}{\begin{equation}}
\newcommand{\ee}{\end{equation}}
\newcommand{\ba}{\begin{eqnarray}}
\newcommand{\ea}{\end{eqnarray}}

\title[Evolution of the Dark Matter Distribution with 3-D Weak
Lensing] {Evolution of the Dark Matter Distribution with 3-D Weak
Lensing} \author[D. J. Bacon et al] {D. J. Bacon$^{1*}$,
A. N. Taylor$^{1}$, M. L. Brown$^{1}$, M. E. Gray$^{2}$, C. Wolf$^3$,
\newauthor K. Meisenheimer$^4$, S. Dye$^5$, L. Wisotzki$^6$,
A. Borch$^4$, M. Kleinheinrich$^4$ \\ $^1$Institute for Astronomy,
Royal Observatory Edinburgh, Blackford Hill, Edinburgh, EH9 3HJ,
U. K.\\ $^2$ School of Physics and Astronomy, The University of
Nottingham, University Park, Nottingham, NG7 2RD, U.K.\\ $^3$Department of
Physics, University of Oxford, Keble Road, Oxford OX1 3RH, U.K.\\
$^4$Max-Planck-Institut f\"{u}r Astronomie, K\"{o}nigstuhl 17,
D-69117, Heidelberg, Germany\\ $^5$Astrophysics Group, Blackett
Laboratory, Imperial College, Prince Consort Road, London SW7 2BW,
U.K. \\ $^6$ Astrophysikalisches Institut Potsdam, An der Sternwarte
16, D-14482 Potsdam, Germany\\ $^*$email: djb@roe.ac.uk} \date{}

\pagerange{\pageref{firstpage}--\pageref{lastpage}}
\pubyear{2003}

\begin{document}

\maketitle

\label{firstpage}

\begin{abstract}
We present a direct detection of the growth of large-scale structure,
using weak gravitational lensing and photometric redshift data from
the COMBO-17 survey. We use deep $R$-band imaging of two
$0.5\times0.5$ square degree fields, affording shear estimates for
over 52000 galaxies; we combine these with photometric redshift
estimates from our 17 band survey, in order to obtain a 3-D shear
field. We find theoretical models for evolving matter power spectra
and correlation functions, and fit the corresponding shear correlation
functions to the data as a function of redshift. We detect the
evolution of the power at the $7.7\sigma$ level given minimal priors,
and measure the rate of evolution for $0<z<1$.  We also fit
correlation functions to our 3-D data as a function of cosmological
parameters $\sigma_8$ and $\Omega_\Lambda$. We find joint constraints
on $\Omega_\Lambda$ and $\sigma_8$, demonstrating an improvement in
accuracy by a factor of 2 over that available from 2D weak lensing for
the same area.
\end{abstract}


\section{Introduction}

The evolution of the distribution of matter is a topic of central
concern to cosmology. We seek to understand the history of matter
environments at all epochs, from the near uniform distribution of
matter at $z\simeq 1000$, to the highly skewed distribution at
present. Detailed knowledge of this evolution will lead to precise
measurements of cosmological parameters such as the dark energy
density $\Omega_v$, and its equation of state $w$
(e.g. Refregier et al 2003, Benabed \& van Waerbeke 2003). Furthermore, a
comparison of the evolution of dark matter and baryonic matter will
afford a deep understanding of the formation of baryonic structures.

There are several approaches available for studying matter
evolution. For example, one can compare the matter power spectrum
inferred from the Cosmic Microwave Background with that inferred from
low-redshift galaxy surveys (e.g. Percival et al 2002, Spergel et al
2003); or if one is principally interested in baryonic structure
evolution, one can survey galaxy counts out to higher redshifts and
directly observe evolution in the count power spectra (eg Phleps \&
Meisenheimer 2003). Here, we will study another approach: direct
measurements of matter power spectrum evolution via weak gravitational
lensing combined with photometric redshifts.

Weak gravitational lensing has proved to be a highly useful method of
obtaining cosmological information. The weak lensing phenomenon is
observed as the slight alignment of neighbouring galaxy images, due to
light rays being deflected by gravitational potential fluctuations
along their light paths. As the deflection is due to the gravitational
environment through which the light passes, the phenomenon is
sensitive to all matter, both dark and baryonic. Lensing therefore
allows measurement of the distribution of the entire matter content of
the universe.

The use of weak lensing (often without redshift estimates for
individual galaxies) to constrain the matter power spectrum is now
well developed. The shear estimates for large numbers of galaxies have
been used to measure the statistics of an effective 2-dimensional
shear field; theoretical calculations of these statistics can be fit
to the data in order to estimate cosmological parameters (see e.g. van
Waerbeke et al 2001, Hoekstra et al 2002, Bacon et al 2003, Brown et
al 2003, Jarvis et al 2003) or the statistics can be used to measure
the dark matter power spectrum (e.g. Pen et al 2003, who measure the
3-D power spectrum at a fixed redshift).

Now, with the increase in availability of reliable photometric
redshifts, there is a great deal of interest in combining redshifts
with weak lensing. For example, Wittman et al (2001, 2003) have used
shear and redshifts to measure the mass and redshift of two clusters
directly from the 3-D shear field. In addition, several recent
theoretical studies examine how to directly reconstruct the full 3-D
gravitational potential from lensing together with photometric
redshifts (Taylor 2001; Hu \& Keeton 2003; Bacon \& Taylor 2003); this
will allow examination of peak statistics to obtain cosmological
parameters, and studies of galaxy formation as a function of environment.

A more directly statistical approach to the shear estimators is to use
redshifts to permit shear power spectrum tomography (e.g. Seljak 1998,
Hu 1999, 2002, Huterer 2002, King \& Schneider 2002b, Heavens
2003). In this methodology, the galaxies are divided into several
redshift bins, in each of which shear correlation functions or power
spectra are measured; by comparing these to theoretical models,
cosmological parameters can be estimated. This paper will utilise an
extension of this approach. This methodology also forms the starting
point for removal of systematic pollutants of the cosmological shear
signal such as intrinsic galaxy alignments (Heymans et al 2004,
Heymans \& Heavens 2003, King \& Schneider 2002a,b). An alternative
3-D statistical method (particularly for measuring $w$ and $w'$) is
provided by Jain \& Taylor (2003), which is further investigated by
Bernstein \& Jain (2004).

In this paper, we seek to measure directly the evolution of the matter
power spectrum. We will achieve this using a maximum likelihood fit
between data and theoretical power spectra evolving with redshift. We
will also fit evolving power spectra expected for various values of
cosmological parameters, in order to demonstrate the power of this
method for precise determinations of dark energy parameters with
future large lensing surveys.

We will use the COMBO-17 survey (Wolf et al 2001) to measure the dark
matter evolution. This survey has already been extensively studied
from a weak shear perspective; in particular, it has yielded precise
measurements of mass-to-light correlations for a supercluster (Gray et
al 2002), measurements of the shear power spectrum arising from
large-scale structure (Brown et al 2003), and 3-D gravitational
potential maps (Taylor et al 2004). One of the great assets of the
COMBO-17 survey is the existence of accurate photometric redshifts
($\Delta z \simeq 0.05$) for galaxies with $z\la 1$. Our previous
statistical lensing study with COMBO-17 only used these redshifts
to obtain median redshifts for lens and source planes; in this study,
we will use the redshifts for each individual galaxy to compare each
galaxy pair's distortion with what would be expected theoretically for
galaxies at these 3-D positions.

The paper is organised as follows. In section 2, we will describe the
construction of the theoretical power spectra which we require. In
particular, we discuss the modeling of the underlying evolving matter
power spectrum; we go on to discuss the shear power spectra expected
given this matter power spectrum. This includes not only power spectra
for shear in a redshift shell, but also for cross power spectra for
shears at two different redshifts. We also find inversion equations
for moving between shear and matter power spectra, and construct
phenomenological models to measure evolution of the power
spectra.

We proceed in section 3 to describe the COMBO-17 survey, giving
details regarding the observations, photometric redshift estimation
and weak shear analysis.

Section 4 describes the analysis of the data. We measure the rate of
evolution of the matter power spectrum using a simple model. We also
find constraints upon the cosmological constant and amplitude of the
power spectrum by fitting models with varying values of these
parameters to the 3-D shear field data.

Finally, in section 5 we discuss the implications of our results,
and discuss the future development of the method on large shear
datasets.

\section{Power Spectra and Correlation Functions}

In this section we will describe our approach to calculating
theoretical models for the matter power spectrum and shear power
spectrum for various cosmologies. We will use these models to compare
with the data, in order to measure cosmological parameters. We also
describe here how we obtain cross-correlation models for shear
estimators at different redshifts, and how one can calculate the
matter power spectrum from the shear power spectrum and vice
versa. Finally, we will present a phenomenological model suited
to a direct measurement of the evolution of the power spectrum.

\subsection{Matter Power Spectrum}

We begin by describing the method used for calculating theoretical
matter power spectra for different values of cosmological
parameters. This will require us to calculate a full non-linear power
spectrum, which we will obtain via the intermediate step of
calculating the linear power spectrum.  First then, we require a model
for the linearly evolving power spectrum $\Delta^2(k)$ in logarithmic
increments of comoving wavenumber $k$. In order to achieve this, we
use an initially Harrison-Zeldovich power spectrum, evolved with a
transfer function $T$, growth factor $g$ described below, and a
$(1+z)^{-2}$ factor as expected for Einstein-de Sitter linear growth
of a matter-dominated expansion (c.f. Bardeen et al 1986):

\begin{equation}
\label{powerspec}
\Delta^2(k,z)=A k^3 T^2(k) k \frac {g^2(z)}{(1+z)^2 g^2(0)},
\end{equation}
where $A$ is a numerical constant which we will normalise presently.
The transfer function $T$ for the Cold Dark Matter scenario is given
by (e.g. Bardeen et al 1986; c.f. Efstathiou, Bond \& White 1992, Bond
\& Efstathiou 1984):

\begin{eqnarray}
T_{{\rm CDM}}(k) & = & \frac{\ln \left(1+2.34q \right)}{2.34q} \times  
	\\ \nonumber 
& & \hspace*{-1.5cm}
	\left[1+3.89q+(16.1q)^2+(5.46q)^3+(6.71q)^4\right]^{-1/4},
\end{eqnarray}
with $q = k/h\Gamma$. Here, $h$ is the present-day Hubble constant
$H_0$ divided by 100 km s$^{-1}$ Mpc$^{-1}$, and the `shape parameter'
$\Gamma$ is defined as (Sugiyama 1995)

\begin{equation}
\label{gamma}
\Gamma = \Omega_{m,0} h \exp (-\Omega_{{\rm B},0}(1-\sqrt{2
h}/\Omega_{m,0})) \, .
\end{equation}
This equation introduces the present-day density parameters for
matter, $\Omega_{m,0}$, and baryons, $\Omega_{B,0}$. In order to
calculate $\Delta^2$, we define a growth parameter $g$ which describes
the difference in growth between a particular cosmology and an
Einstein-de Sitter cosmology at a given redshift $z$ (Carroll et al
1992):

\begin{eqnarray}
\label{supp}
g(z) = \frac{5}{2} \Omega_m(z)
\left[\Omega_m^{4/7}(z)-\Omega_v(z)+\left(1+\frac{1}{2}
\Omega_m(z)\right)\right. \\
\left.\left(1+\frac{1}{70}\Omega_v(z)\right)\right]^{-1} .
\end{eqnarray}
In turn this equation requires the evolution equations for the density
parameters of matter, $\Omega_m$, and of the vacuum energy,
$\Omega_v$,

\begin{equation}
\Omega_m(z)=\frac{\Omega_{m,0}} {a + (1-a) \Omega_{m,0} + (a^3-a) \Omega_{v,0}}
\end{equation}

\begin{equation}
\Omega_v(z)=\frac{a^3 \Omega_{v,0}}{a + (1-a) \Omega_{m,0} + (a^3-a) \Omega_{v,0}},
\end{equation}
where the expansion parameter $a=1/(1+z)$. Now we have the equipment
for calculating the linear power spectrum, we can progress to
estimating the non-linear evolution of the power spectrum, via the
formalism of Smith et al (2003). 

This method calculates the fully non-linear power using the insights afforded
by the Halo Model (e.g. Ma \& Fry 2000, Peacock \& Smith 2000, Seljak
2000). The power is decomposed into two terms, a quasi-linear term
describing the power from the clustering of halos, and a halo term
describing power from the clustering of objects within each halo. Each
of these terms is calculated by Gaussian filtering the linear power to
find a non-linear threshold wavenumber $k_\sigma$, spectral index and
spectral curvature. These are used to find 11 coefficients which
are functions of the calculated spectral properties; the quasi-linear and halo
power terms are then found as functions of the linear power and these
coefficients, finally resulting in the full power $\Delta^2$. Full
details of this procedure can be found in Smith et al (2003).

Finally we move from the dimensionless power $\Delta^2$ to the standard
power spectrum $P$ via

\begin{equation}
P_{NL}(k,z)=2\pi^2 \Delta^2_{NL}(k,z)/ k^3 .
\end{equation}

We normalise the power spectrum at the present epoch by setting a value
for the quantity $\sigma_8$, measuring the rms fluctuation of density
when smoothed by a top-hat filter with radius $R=8h^{-1}$Mpc:

\begin{equation}
\sigma_8^2 = \int_0^\infty k^2 dk P(k,0) \frac{3}{(k R)^3} (\sin (k R) -
k R \cos (k R)) .
\end{equation}

Thus, for particular choices of $\sigma_8, \Omega_{m,0}, \Omega_{v,0},
h, \Omega_b$ and $\Omega_k$, we have a model for the evolving non-linear power
spectrum at any required redshift. 

\subsection{Shear Power Spectrum}

Now that we have a model for the matter power spectrum at any epoch,
we can proceed to calculate the 3-D shear power spectrum and
correlation function.  It is convenient to obtain these via the
effective convergence $\kappa$, a measure of the surface mass density,
which is given by (following Bartelmann \& Schneider 2001, Ch 6
throughout this section):

\begin{equation}
  \kappa({\mathbf \theta}) =
  \frac{3H_0^2\Omega_m}{2c^2}\,
  \int_0^{r_{\mathrm H}}\,d r\,W(r)\,f_K(r)\,
  \frac{\delta[f_K(r) {\mathbf \theta},r]}{a(r)}\;.
\label{eq:kappa}
\end{equation}
Here, $\theta$ represents the position on an image, $\delta$ is the
density perturbation field, $c$ is the speed of light, $r_H$ is the
comoving radial distance to the horizon, and $r$ is the comoving
radial distance to a given mass concentration:

\begin{equation}
r(z)=\frac{c}{H_0}\int_{a(z)}^1 [a \Omega_m(0) + a^2
(1-\Omega_m(0)-\Omega_v(0))+a^4 \Omega_v(0)]^{-1/2} da .
\end{equation}

In the equation for $\kappa$ above we also require the radial
coordinate distance $f_K$,

\begin{equation}
  f_K(r) = \left\{
  \begin{array}{ll}
    K^{-1/2}\sin(K^{1/2}r) & (K>0) \\
    r & (K=0) \\
    (-K)^{-1/2}\sinh[(-K)^{1/2}r] & (K<0) \\
  \end{array}\right.\;
\label{eq:fk}
\end{equation}
with the curvature $K$ given by

\begin{equation}
K=\left(\frac{H_0}{c}\right)^2(\Omega_m+\Omega_v-1) .
\end{equation}

We also require the weighting factor $W$,

\begin{equation}
W(r) = \frac{f_K(r_s-r)}{f_K(r_s)} H(r_s-r) ,
\end{equation}
where $H$ is the Heaviside function and $r_s$ is the comoving distance
to a particular source galaxy. 

We can now find the cross-power between convergence at any two
redshifts. In order to do this we use Limber's equation, which can be
formulated (e.g. Bartelmann \& Schneider 2001) to state that if we
have two projections $p_i, i=1,2$ of the density field $\delta$, such
that

\begin{equation}
  p_i({\mathbf \theta}) = \int d r\,q_i(r)\,\delta[f_K(r){\mathbf \theta},r]\;,
\label{eq:pi}
\end{equation}
for some function $q_i(r)$, then the cross power spectrum will be

\begin{equation}
  P_{12}(\ell) = \int d r\,\frac{q_1(r)q_2(r)}{f_K^2(r)}\,
  P_\delta \left( \frac{\ell}{f_K(r)}, r\right)\; .
\end{equation}

If we wish to calculate the cross power of $\kappa$ between two
redshifts, then we see by comparing equations (\ref{eq:kappa}) and
(\ref{eq:pi}) that we do have two projections of the necessary form,
with $q_i$ given by

\begin{equation}
  q_i(r)=\frac{3}{2}\,\frac{H_0^2}{c^2}\,\Omega_m\,\frac{f_K(r)
  f_K(r_i-r)}{f_K(r_i)} \frac{H(r_i-r)}{a(r)} ,
\end{equation}
so we therefore obtain (noting that the two-point statistics of
$\kappa$ and $\gamma$ agree, e.g. Blandford et al 1991),

\begin{eqnarray}
  P_{\gamma 1 2}(\ell,z_1,z_2) = \frac{9H_0^4\Omega_m^2}{4c^4}\,
  \int_0^{r_1 < r_2}\, d r\,
  \frac{f_K(r_1-r)}{f_K(r_1)}\frac{f_K(r_2-r)}{f_K(r_2)} \nonumber \\
  \frac{1}{a^2(r)}P_\delta\left(\frac{\ell}{f_K(r)},r\right) ,
\label{eq:pg2}
\end{eqnarray}
where $\ell$ is the angular wavenumber. In the case where we wish to
examine the power of the shear at one particular redshift, we can
simplify this to

\begin{equation}
  P_{\gamma}(\ell,z_1) = \frac{9H_0^4\Omega_m^2}{4c^4}\,
  \int_0^{r_1}\, d r\,
  \frac{f_K^2(r_1-r)}{f_K^2(r_1) a^2(r)}
  P_\delta\left(\frac{\ell}{f_K(r)},r\right) .
\label{eq:pg}
\end{equation}

We can obtain corresponding cross-correlation functions for shears at
two different redshifts $z_1$, $z_2$. We will use three different
correlation functions: $C_1$ represents $\langle \gamma_1^a \gamma_1^b
\rangle$, where $\gamma_1^{a,b}$ represents the first shear component
of two galaxies, in a coordinate frame where zero position angle lies
along the line joining the galaxies (e.g. Bacon et al
2003). $C_2$ represents $\langle \gamma_2^a \gamma_2^b \rangle$, and
$C=C_1+C_2$. Given these definitions, the correlation functions are
simple transforms of the power:

\begin{equation}
C(\theta,z1,z2)=\int_0^\infty \frac{\ell d\ell}{2 \pi} P_{\gamma 1 2}(\ell,z_1,z_2)
J_0(\ell \theta) ,
\label{ctheta}
\end{equation}

\begin{equation}
C_1(\theta,z1,z2)=\int_0^\infty \frac{\ell d\ell}{4 \pi} P_{\gamma 1 2}(\ell,z_1,z_2)
[J_0(\ell \theta)+J_4(\ell \theta)] ,
\label{ctheta1}
\end{equation}

\begin{equation}
C_2(\theta,z1,z2)=\int_0^\infty \frac{\ell d\ell}{4 \pi} P_{\gamma 1 2}(\ell,z_1,z_2)
[J_0(\ell \theta)-J_4(\ell \theta)] .
\label{ctheta2}
\end{equation}

These are the 3-D shear correlation functions we
have been seeking; we will use these functions to compare our data
with various evolving cosmological models.

\subsection{Power Spectrum Inversion}

It is convenient to have a simple means of calculating the shear power
spectrum from the matter power spectrum; this has been described in
the last section. However, it is also useful to have a means of
calculating the matter power spectrum given the shear power spectrum;
here we outline how this can be achieved for theoretical models. Noisy
data are difficult to invert with this procedure; it would be more
convenient to fit models to the data and then use the results of this
section to invert these models.

In order to show how to invert the shear power spectrum, we
start by considering an integral of the form

\begin{equation}
A(r)\equiv \int_0^r dr' B(r',r) ,
\end{equation}
where $B(r',r)$ is a smooth, continuous function. Carrying out a partial
differentiation of $A$ with respect to $r$, we find

\begin{equation}
\frac{\partial A(r)}{\partial r}=\int_0^r dr' \frac{\partial
B(r,r')}{\partial r}+B(r,r) .
\end{equation}
This result will be used below in order to untangle the matter power
spectrum from its integral projection found in calculating the
shear. We also require the result

\begin{equation}
\frac{\partial}{\partial r_1}
\left(\frac{f_K(r_1-r)}{f_K(r_1)}\right)=\frac{f_K(r)}{f_K^2(r_1)} ,
\end{equation}
which can be easily verified directly from equation
(\ref{eq:fk}). Using these two results repeatedly upon the shear power
spectrum $P_{\gamma}(l,r)$, we find that we can calculate the matter
power spectrum:

\begin{eqnarray}
P_\delta(k,r_1)=\frac{4 c^4}{9 H_0^4 \Omega_m^2}\frac{a^2(r_1)}{2 f_K(r_1)}
\left(\frac{\partial^2}{\partial r_1^2}+K \right)\nonumber\\ \left[
f_K^3(r_1)\frac{\partial P_\gamma(f_K(r_1)k,r_1)}{\partial r_1}\right] .
\label{eq:invert}
\end{eqnarray}

Alternatively, we can find a similar means of calculating the matter
power spectrum from the cross power spectrum of shear at two different
redshifts: 

\begin{eqnarray}
P_\delta(k,r_1)=\frac{4 c^4}{9 H_0^4
\Omega_m^2}\frac{a^2(r_1)f_K(r_2)}{f_K(r_1)
f_K(r_2-r_1)}\frac{\partial}{\partial r_1}\nonumber\\
\left(f_K^2(r_1)\frac{\partial P_{\gamma 1
2}(f_K(r_1)k,r_1,r_2)}{\partial r_1}\right) .
\label{eq:invert2}
\end{eqnarray}

These equations therefore permit us to find the matter power spectrum
given a model for the shear power spectrum. Note that upon noisy data,
the first and second differentials in this equation can lead to
unphysical negative power spectra. Thus we reiterate that a better
approach is to fit a continuous shear model to the data which can then
be inverted with these equations.

\subsection{Power Spectrum Model for Slope Phenomenology}

We conclude this section with a discussion of models for the
matter and shear power spectra which allow us to directly examine the
redshift evolution of the power spectrum.

Firstly, suppose we have measured the shear cross-correlation function
$C$ between many redshifts, from which we can calculate the shear
power spectrum. We will initially restrict ourselves to power within a
redshift shell, $P_\gamma$; we will consider $P_{\gamma 1 2}$
later. We will attempt to use our formalism with a simple model for
the shear power spectrum: a power law in both the angular and redshift
directions, i.e.

\begin{equation}
P_{\gamma}(l,r) = A l^\alpha r^\beta = A k^\alpha
r^{\alpha+\beta} .
\label{eq:powlawspec}
\end{equation}

The correlation function corresponding to this model, calculated from
equation (\ref{ctheta}) is:

\begin{equation}
C(\theta,r)=\frac{A 2^{\alpha-1} \alpha \Gamma(\alpha/2)}{\pi
\Gamma(-\alpha/2)} \theta^{-2-\alpha} r^\beta .
\end{equation}
(One should note more generally that, if $P_\gamma\propto k^\alpha
f(r)$, then $C\propto \theta^{-2-\alpha} r^{-\alpha}f(r)$.)

Hence, if we fit a power law to the correlation function in the
angular and redshift directions, we can immediately obtain estimates
of the corresponding power-law approximation to the shear power
spectrum, and an approximation to the underlying matter power
spectrum. To achieve the latter, we use the inversion equation
(\ref{eq:invert}); we find that the matter power spectrum is given (in
a flat universe) by

\begin{equation}
P_\delta(k,r)=\frac{4 c^4}{9 H_0^4 \Omega_m^2} A (\alpha+\beta)
(\alpha+\beta+1) (\alpha+\beta+2) k^\alpha a^2(r) r^{\alpha +\beta-1} .
\end{equation}

This is a very useful formula; if we can adequately fit the shear
power spectrum as a power law in the angular and redshift directions,
we can instantly infer the corresponding power law matter power
spectrum.

We can attempt a similar approach with the cross power spectrum,
$P_{\gamma 1 2}$, with a power law model keeping one redshift fixed:

\begin{equation}
P_{\gamma 1 2}(l,r_1, r_2) = A l^\alpha r_1^\beta = A k^\alpha
r_1^{\alpha+\beta} .
\end{equation}

However, this is less successful, as we obtain the following form for
the matter power spectrum:

\begin{equation}
P_\delta(k,r_1)=\frac{4 c^4}{9 H_0^4 \Omega_m^2} A (\alpha+\beta) (\alpha+\beta+1) r_2 k^\alpha \frac{a^2(r_1)
r_1^{\alpha +\beta-1}}{r_2-r_1} .
\end{equation}

This model fails to be physical; besides an unsightly divergence at
$r_2=r_1$, we find that the matter power at the present epoch is
either zero or infinite, depending on the power law slope. We
therefore conclude that a power law model for the cross shear power is
unsuitable for directly calculating the matter power
spectrum. Equivalently, we should note that the underlying matter
power spectrum which produces a power-law cross shear power spectrum is
unphysical.

In order to find a measure of matter power spectrum evolution, able to
take into account all the information from the shear cross power spectrum,
one can take an alternative approach. We will assume a very simple
form for the underlying matter power spectrum,

\begin{equation}
P_\delta(k,r_1)=\frac{4 c^4}{9 H_0^4 \Omega_m^2} A k^\alpha {\rm
e}^{-sz} ,
\label{pdelta}
\end{equation}
and will restrict ourselves to an underlying $\Lambda$CDM geometry in
moving from this matter power to the shear predictions. This allows us
to directly examine the matter evolution, rather than examining both
matter and geometry change. The form chosen for $P_\delta$ has the
advantage of being positive definite as required for power; it is also
a good fit to $\Lambda$CDM evolution in the non-linear regime that we
wish to study.

In order to remove some of the freedom from the above equation, we
note that $\alpha=-1.2$ provides an excellent fit to the angular
dependence of the $\Lambda$CDM shear correlation function throughout
$z<1$ and $1'<\theta<30'$.  We therefore allow only the matter
amplitude $A$ and redshift evolution slope $s$ to vary. We will fit
the corresponding shear correlation function calculated from equation
(\ref{eq:pg2}) to the data, allowing us to directly measure the rate
of evolution of the power spectrum.

\section{Data}

Having developed a suitable formalism for examining the 3-D matter
power spectrum, we wish to proceed to measure this directly from the
COMBO-17 survey (Wolf et al 2001). This survey currently spans 1
square degree, in four widely separated $0.5\times0.5$ square degree
fields. One of the fields contains the A901/2 supercluster, and is
therefore deemed highly unrepresentative and inappropriate for
inclusion in this study (c.f. the significantly anomalous power
spectrum measured for this field in Brown et al 2003). One further
field (SGP) does not currently have full redshift information
available, so this study focuses upon the two remaining fields,
centred upon the Chandra Deep Field South (CDFS) and an area with no
previously known mass concentrations (S11), but which in reality
contains a cluster at $z=0.11$; Brown et al (2003) show that this does
not produce a high power spectrum estimate in this field.

All fields in the survey were observed at the MPG/ESO 2.2m telescope,
La Silla, Chile, with the Wide-Field Imager. This is a $4\times2$
array of $2048\times 4096$ pixel CCDs, with pixel scale 0.238
arcseconds (c.f. Gray et al 2002).

The key characteristic of the COMBO-17 data which makes it suitable
for our purpose is the existence of an extensive catalogue of accurate
photometric redshifts for the galaxies in each field. Each COMBO-17
field was observed in 5 broad-band filters ($UBVRI$) and 12
medium-band filters running from 350nm to 930nm. This permits
simultaneous estimates of redshift and SED classification using
empirically-based spectral templates; this is described in detail in
Wolf et al (2001), who use the same version of the catalogue. The
resulting accuracy of photometric redshift is $\Delta_z\simeq0.05$ for
galaxies with $z<1.0$.

The weak shear measurements were made upon the $R$ filter images from
the survey. This filter was used in the best seeing conditions,
providing a deep image optimal for weak lensing; the combined exposure
length was 36120 seconds for the CDFS field, and 18100 seconds for the
S11 field.

The reduction process for this $R$ band imaging data is described in
Gray et al (2002) and Brown et al (2003). For the CDFS field, $78\times8$ CCD
chip exposures were registered by using linear astrometric fits,
including a 3$\sigma$ rejection of bad pixels and columns; for the S11
field, $44\times8$ exposures were registered in this fashion.

The shear estimation for the galaxies was carried out using the {\tt
imcat} package, following the methodology of Kaiser, Squires \&
Broadhurst (1995); see Gray et al (2002) and Brown et al (2003) for
details of the implementation for our dataset. The analysis yields a
catalogue of galaxies with centroid positions, plus shear estimates
which have been corrected for circularisation of the galaxies by the
PSF, and anisotropic smearing due to e.g. tracking errors. This
catalogue was combined with the photometric redshift catalogue for the
COMBO-17 survey. 

The resulting catalogue included 52139 galaxies, 23102 of which had
reliable photometric redshifts assigned; the remaining galaxies
yielded ambiguous redshift probabilities or were fainter than the
$R=24$ reliability limit for redshifts in the survey. This remainder
of galaxies was flagged as having unreliable redshifts; the faint
galaxies were assigned an optional redshift which we use below when
stated. Following Brown et al (2003), Section 4, we use a
median-magnitude median-redshift relation to assign this redshift; the
median $R$ magnitude of the faint sample is $R=24.9$, corresponding to
a median redshift $z\simeq 0.95\pm0.05$.

\section{Analysis}

We will now use the models developed above to assess the evidence for
evolution of the matter distribution from the COMBO-17 survey. We begin
by making a minimal direct detection of evolution, by comparing the
growth of the lensing signal with redshift to that expected with a
truly unevolving matter distribution. We go on to use our
phenomenological model of an evolving power spectrum to constrain the
rate of evolution. Finally, we return to full power spectra models
including cosmological parameters, in order to constrain the
cosmological constant from the rate of growth of the shear signal with
redshift.

\subsection{Preliminary detection of evolution}

As a preliminary first step in using 3-D shear data to measure
evolution, we will demonstrate that the data exclude a non-evolving
matter distribution, given the present-day normalisation of the matter
power spectrum $\sigma_{8,0}$ and a $\Lambda$CDM geometry.  It is
important to realise that this simplest step is already effectively
achieved by 2D cosmic shear surveys, if the median redshift of the
survey is well estimated. That is to say, the amplitude of the
projected 2D shear field correlation functions measured with cosmic
shear surveys (see e.g. van Waerbeke \& Mellier 2003, Refregier 2003
for reviews) at a median redshift $z_m \sim 1$ is inconsistent with
that expected for an unevolving power spectrum, both calculated for
example with $\sigma_{8,0}=0.84\pm0.04$ (c.f. Spergel et al 2003).
For example, at an angular scale of $1'$, $C_0\simeq
(3.2\pm0.4)\times10^{-4}$ for $z_m=1$ in these surveys, whereas $C_0$
is predicted to be $(5.1\pm0.8)\times10^{-4}$ at 1' in a model with
unevolving power (i.e. with the power expected for $\Lambda$CDM at the
present day as in section 2.1, but unchanged in the past),
$\sigma_{8,0}=0.84\pm0.04$ and $\Lambda$CDM geometry $(\Omega_m=0.3,
\Omega_{\rm tot}=1, H_0=72$km s$^{-1}$ Mpc$^{-1}$) calculated using
equation (\ref{eq:pg}).

However, we will see that there is considerable power in using the
redshift information for each galaxy, as we will obtain significantly
more accurate measurements of the evolution, and will be able to
exclude no-growth models regardless of present-day power spectrum
normalisation. We begin by making a direct detection of evolution with
the 3-D shear field. We can do this by calculating the $\chi^2$ fit of
the no-evolution model to the data, and comparing with the $\chi^2$
fit for, say, an evolving $\Lambda$CDM model $(\Omega_m=0.3,
\Omega_{\rm tot}=1, H_0=72$km s$^{-1}$ Mpc$^{-1}$), in the first
instance both with $\sigma_{8,0}=0.84$ (c.f. Spergel et al 2003). In
the next section we will examine the effect of varying this
normalisation and marginalising over cosmological parameters.

The sum we require for our $\chi^2$ fit is given by

\begin{eqnarray}
\chi^2= \sum_{i,j} w_{ij} [(\gamma_{1,i} \gamma_{1,j} -
C_1(\theta_{ij},z_i,z_j))^2+ \nonumber \\ (\gamma_{2,i} \gamma_{2,j} -
C_2(\theta_{ij},z_i,z_j))^2 ] ,
\label{chi2}
\end{eqnarray}
where $\gamma_1$ and $\gamma_2$ are the two components of shear for a
particular galaxy, in the frame where the $x$-axis is the line joining
the two galaxies in question; $w_{ij}$ are a set of weights which we
will choose presently. $i$ and $j$ are indices for different galaxies,
and $C_1$ and $C_2$ are the expected correlation functions between the
two galaxies in the model, calculated using equations (\ref{ctheta1})
and (\ref{ctheta2}).

Here we choose a simple weighting $w_{ij}=1/\sigma_\gamma^4$, $i \ne
j$, where $\sigma_\gamma$ is the variance in one shear component. We
choose this weighting as the error upon $\gamma_{1,i} \gamma_{1,j}$ is
entirely dominated by the random orientation of galaxies; therefore we
have neglected off-diagonal elements of the covariance matrix of
errors upon $\gamma_{1,i} \gamma_{1,j}$. In addition, we neglect the
smoothing error associated with our redshift uncertainty of $\Delta
z\simeq 0.05$; assessment of the impact of this small smoothing term
is left for future investigation. We are also invoking the Central
Limit Theorem in order to use $\chi^2$ to estimate uncertainties upon
parameters, as any one product of galaxy shears does not have a
Gaussian error.

In order to remove the impact of intrinsic alignments between
galaxies, due to physical galaxy alignment instead of apparent
alignment due to lensing, we reject galaxy pairs which are within
$\Delta z = 0.05$ of each other, if their comoving separation is less
than 1Mpc. We also remove, with the same prescription, near galaxy
pairs in the background without redshift information, assigning to
them the median redshift as described in Section 3.

The weighting equation (\ref{chi2}) is $1/\sigma_\gamma^4$ since,
if our shear estimator $\gamma_i$ is near Gaussian, we find
that the probability distribution of $\gamma_i \gamma_j$ (i.e. two
independent Gaussian variables multiplied together) is given by

\begin{equation}
{\rm Prob}(y=\gamma_i \gamma_j)=\frac{1}{\pi \sigma_\gamma^2}
K_0\left(\frac{y}{\sigma_\gamma^2}\right) ,
\end{equation}
where $K_0$ is a modified Bessel function of the second kind. This
probability distribution, ${\rm Prob}(y)$, has variance
$\sigma_\gamma^4$ as stated; this is found to be an excellent match to
the variance of measured $\gamma_i \gamma_j$ (both have value
$0.31^4$).

The no-evolution model we select is as described above,
i.e. $\Lambda$CDM $(\Omega_m=0.3, \Omega_{\rm tot}=1, H_0=72$km
s$^{-1}$ Mpc$^{-1}$) power at the present day with
$\sigma_{8,0}=0.84$; however, this power remains constant at earlier
epochs. An evolving $\Lambda$CDM geometry is assigned with parameters
as above, evolving as appropriate to a given epoch; that is to say,
the lack of evolution is directly in the power, not in the
geometry. As described in Section 3, we apply the fit to two cases: in
the first case, we include only galaxies with a reliable photometric
redshift measurement. In the second case, we assign a redshift
$z=0.95$ to objects without measured redshift, for both evolving and
non-evolving measurements of $\chi^2$. (In the next section, we will
include the effects of marginalisation over the allowed range of
median redshift and Hubble parameter).

If we include the background objects with assigned median redshift, we
find that the probability ratio between the no-evolution model and the
$\Lambda$CDM model is $\exp{(\chi_{\rm noev}^2-\chi_{\Lambda {\rm
CDM}}^2)/2}={\rm e}^{-31.1}$; thus the no-evolution model is
completely excluded given this simple two-model comparison. If one
only uses objects with known redshifts, one finds a probability ratio
of 49.4; i.e. the $\Lambda$CDM model is approximately 50 times more
likely a model than the no-evolution model, given our data.

Thus we see that, using shear estimators together with photometric
redshifts for each galaxy, we are immediately able to set powerful
constraints on the evolution of structure. But we must now proceed to
a fuller treatment for measuring the evolution, including the effect
of power spectrum normalisation.

\subsection{Measuring the evolution of the power spectrum}

In this section, we seek to measure the slope of the mass power
spectrum in the redshift direction, together with the present-day
amplitude of the power spectrum. In order to achieve this, we use the
parameterised power spectrum of equation (\ref{pdelta}), rescaled as
$P_\delta(k,w)=A k^\alpha {\rm e}^{-s z}$, and calculate from this the
shear correlation functions as described in Section 2.2, using our
standard $\Lambda$CDM geometry $(\Omega_m=0.3, \Omega_{\rm tot}=1,
H_0=72\pm5$km s$^{-1}$ Mpc$^{-1}$ from Freedman et al 2001). We carry
out the $\chi^2$ calculation above (equation (\ref{chi2})), varying
the parameters $A$ (power spectrum amplitude at present day) and $s$
(related to the gradient of the spectrum as a function of redshift),
while fixing $\alpha=-1.2$ as discussed in Section 2. We marginalise
over $H_0=72\pm5$km s$^{-1}$ and median redshift $z\simeq
0.95\pm0.05$.

\begin{figure}
\psfig{figure=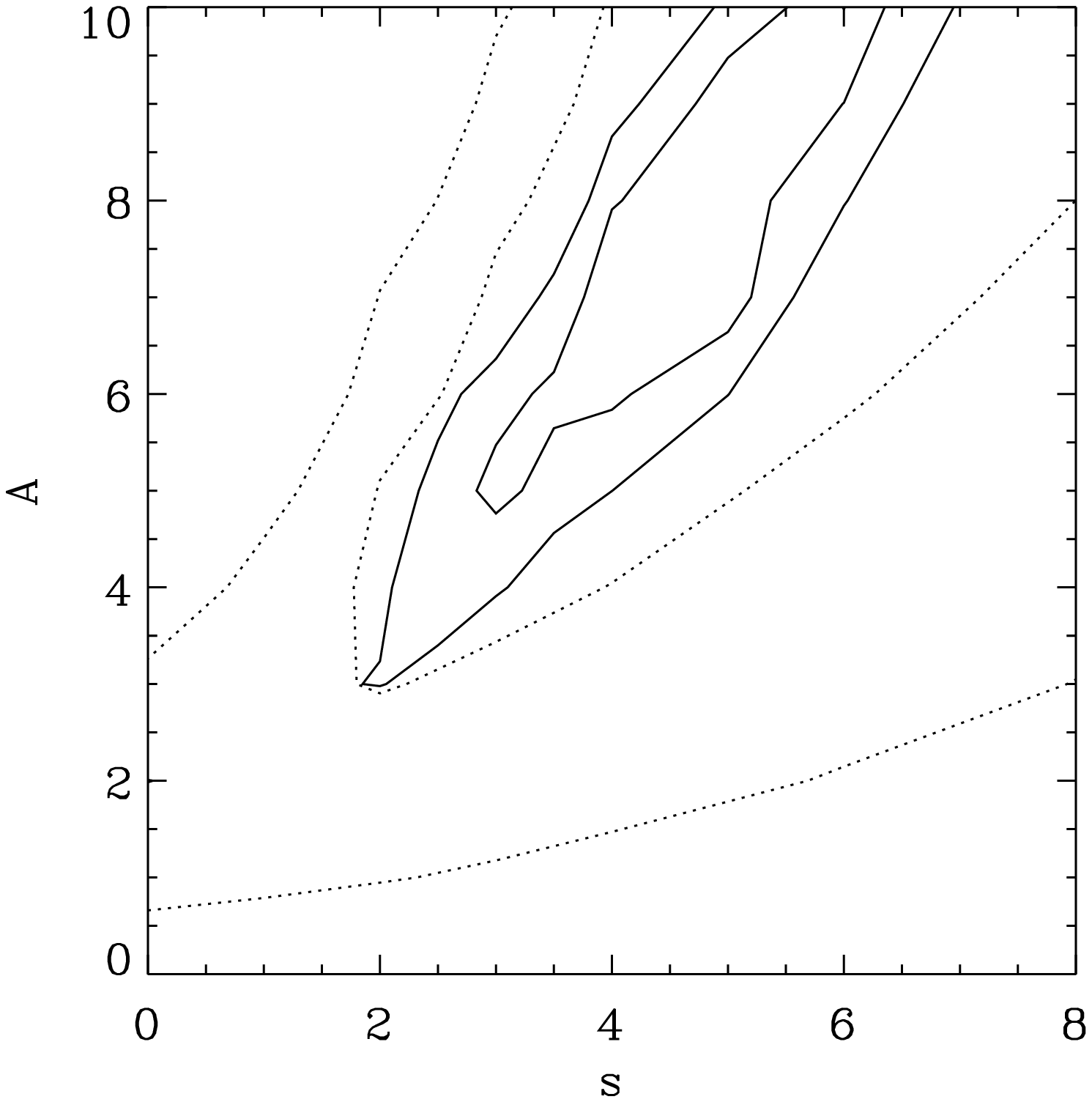,height=7cm,angle=0}
\psfig{figure=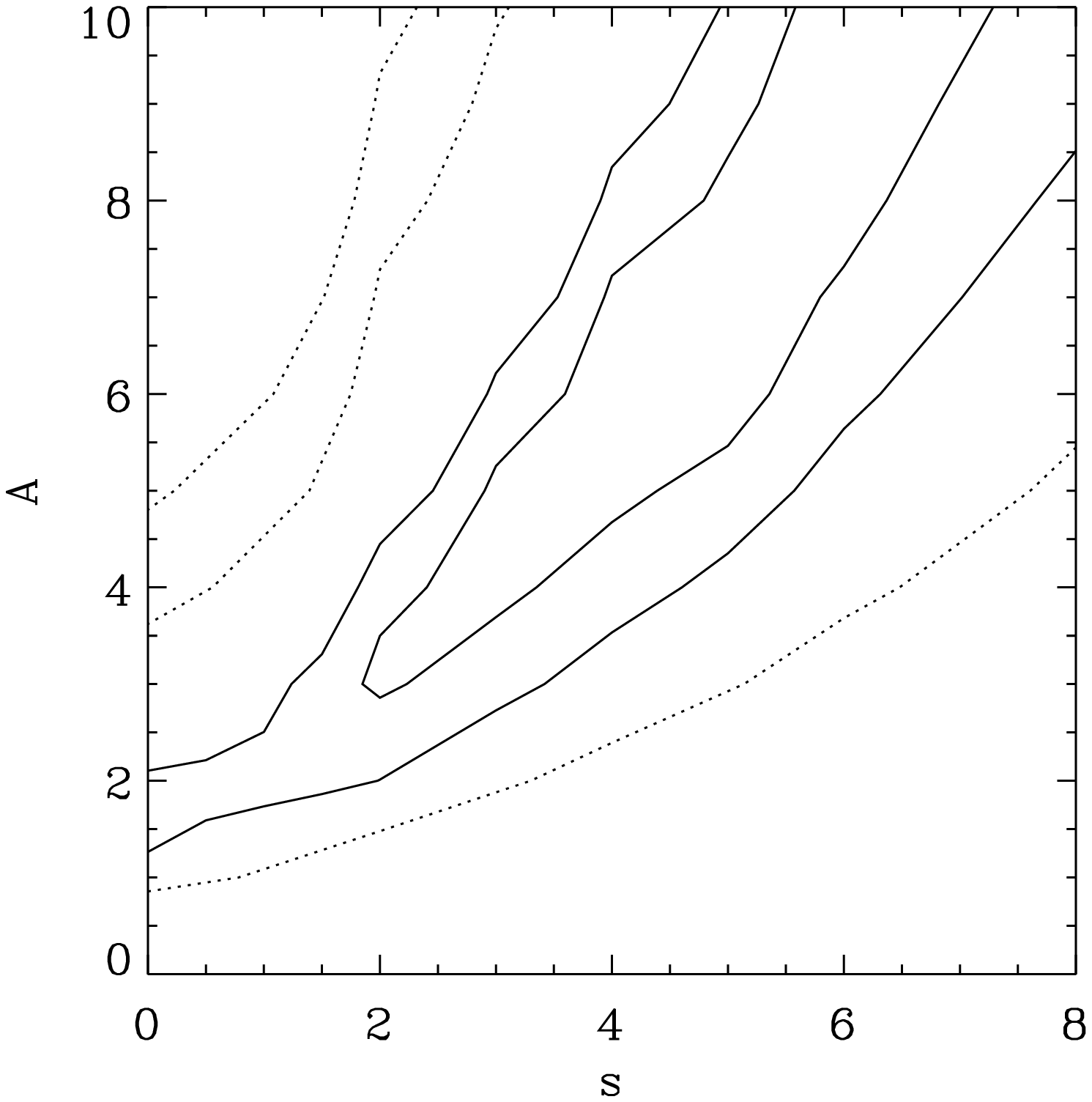,height=7cm,angle=0}
\psfig{figure=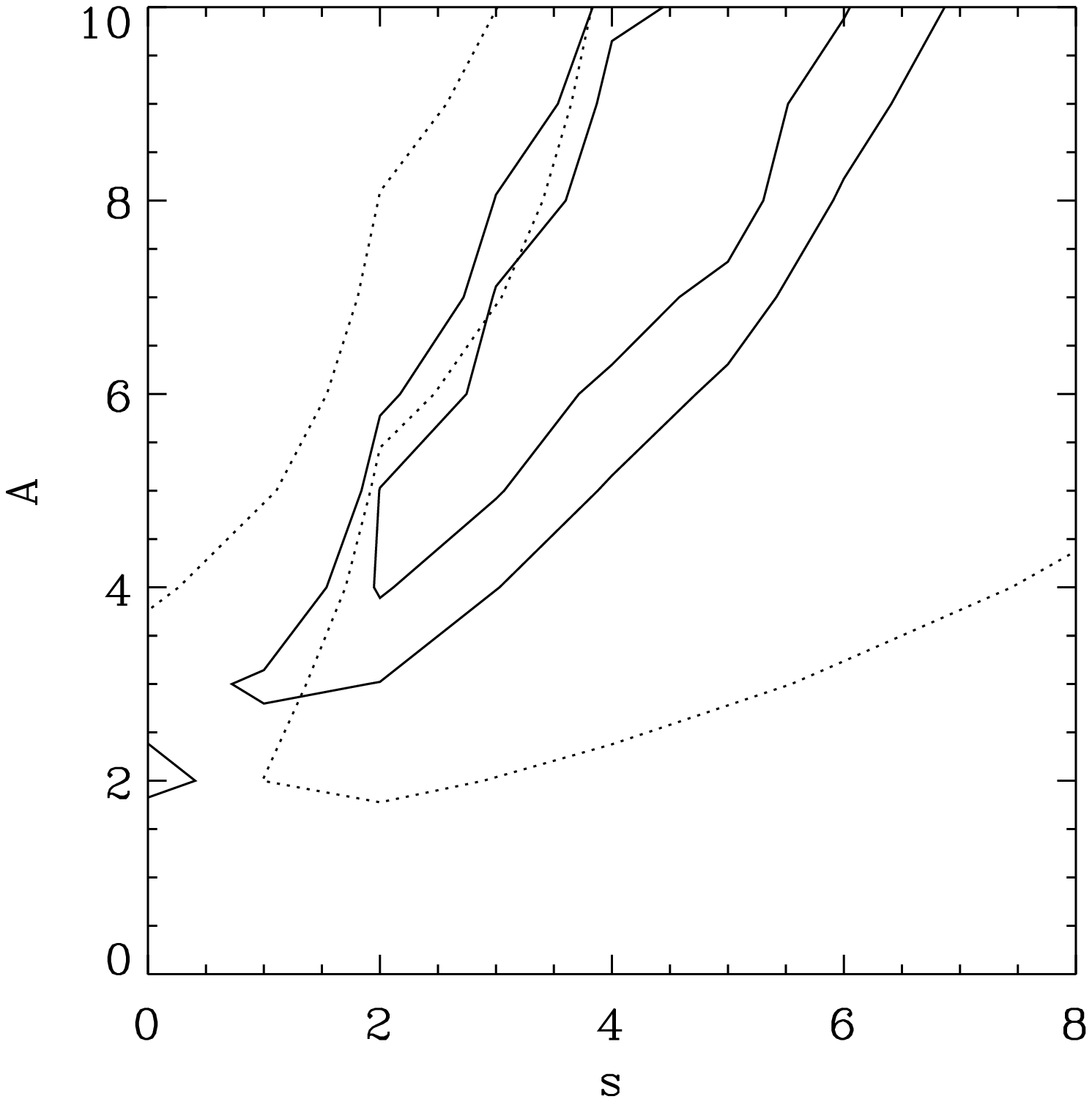,height=7cm,angle=0}
\caption{Joint constraints on power spectrum amplitude $A$ and
evolution slope $s$ from COMBO-17 survey (CDFS and S11 fields). Top
panel: combined constraints from the two fields. Dotted lines show the
constraints from galaxies with known redshifts only. Contours display
1 and 2$\sigma$ confidence regions. Solid lines show constraints
including galaxies with unknown redshift, assigned z=0.95. Middle
panel: constraints from the CDFS field only. Bottom panel: constraints
from the S11 field only.}
\end{figure}

The resulting constraints on the parameters $A$ and $s$ are shown in
Figure 1. The panels show the constraints from the combined fields,
CDFS and S11 respectively; dotted lines show the constraints if we only
include galaxies with known redshifts, while solid lines show
constraints including the galaxies with unknown redshifts.

It is easy to understand the basic shape of the confidence contours;
if one has a higher amplitude of power at the current epoch, one
requires a more dramatic rate of growth to account for the shear
estimates (c.f. the previous section, where we saw that a moderate
present-day amplitude absolutely requires decreasing power as we increase
redshift).

Note that our fit allows very large values of the present-day
normalisation $A$ and slope $s$ (Figure 1 only shows the $A-s$ region
of interest; acceptable fit is achieved even for $A=100$). This is
because the lensing predictions from a finite field are insensitive to
the power and growth of power present at low redshift (c.f. discussion
concerning Figure 2 below). Thus for our investigations, we will
impose a prior $A<10$ corresponding to a rate of growth approximately
three times faster than that expected for $\Lambda$CDM, as discussed
below.

We find that $s<0$ is excluded at the 1.9$\sigma$ level if we include
all galaxies, and at the 1.2$\sigma$ level if we only include galaxies
with known redshifts. This initially appears to be a much lower level
of confidence to that discussed in the previous section; however, here
we are still allowing models which have very low present-day power
spectrum amplitudes. If we require $A>3$, corresponding to
$\sigma_8\ga0.4$, we find that $s\le 0$ is excluded at the $7.7\sigma$
level if all galaxies are included, and at $2.7\sigma$ in the
known-redshift case; therefore we are again finding very strong
rejection of the no-evolution model, if we require a reasonable
present-day power spectrum amplitude.

Turning now to the actual measurements of $A$ and $s$ afforded by our
fit, we find that our 2-D contours constrain $s>2 (1\sigma)$ for CDFS
or S11 independently (including galaxies with unknown redshifts). This
is entirely consistent with our fiducial $\Lambda$CDM model, which is
found to have (by fitting a tangent to the calculated power spectrum
model) an initial ($z=0$) slope coefficient $s=1.7$ for
$k=1$Mpc$^{-1}$ and $s=2.0$ for $k=50$Mpc$^{-1}$. 

The amplitudes of the two fields show significant variation from field
to field, with best fit $A=3.2$ and $A=4.5$ at $s=2$ for CDFS and S11
respectively (including galaxies with unknown redshift); thus when the
$\chi^2$ are combined for the two fields, the reduced overlap
artificially increases the acceptable values of $s$ for the combined
data. Clearly therefore, we will greatly benefit from applying the
methodology to many more lines of sight.

A possible extension of this approach is to directly couch $A$ in
terms of $\sigma_8$; however, since our phenomenological power
spectrum model has a different $k$ and $z$ dependence to a full
cosmologically parameterised power spectrum, values of $\sigma_8$
calculated for our model do not have the conventional
calibration. Instead, we will carry out the equivalent measurement of
the normalisation of the power spectrum in terms of $\sigma_8$ in
section 4.3.

One should note the extraordinary improvement in accuracy arising from
using our whole sample, as opposed to only those galaxies with known
redshifts. The combined constraint upon $A$ and $s$ has improved by a
factor of $>2$ for each field. This acts as a driving concern for
future surveys; we would clearly wish to have a sample with known
redshifts to $R\simeq25$, which poses a serious challenge to current
photometric redshift techniques.

We can use the information we have acquired on $A$ and $s$ to plot
constraints upon the power spectrum growth itself. Since we can
describe $P_\delta=A {\rm e}^{-sz} k^\alpha$, and since we have
measured the probability distribution of $A$ and $s$, ${\rm Prob}(A,s)$,
we can calculate the probability of any particular value of $P_\delta$
at a given $z$, ${\rm Prob}(P_\delta)$.

We calculate this by examining the cumulative probability distribution $F$
for $P_\delta$, i.e. the probability that $P_\delta$ is smaller than a
certain value $P_{\delta 0}$. We see that

\begin{equation}
F=\int_{0}^{P_{\delta 0}} {\rm Prob}(P_\delta) dP_\delta =
\int_{0}^{\infty} dA
\int_{\frac{1}{z}\log{\left(\frac{A}{P_\delta}\right)}}^{\infty} ds
{\rm Prob}(A,s) .
\end{equation}

Now ${\rm Prob}(P_\delta)=\partial F / \partial P_\delta$, so

\begin{equation}
{\rm Prob}(P_\delta)=\int_{0}^{\infty} \frac{dA}{P_\delta z} {\rm
Prob}\left(A,s=\frac{1}{z}\log{\left(\frac{A}{P_\delta}\right)}\right) .
\end{equation}

We can calculate this directly from our $(A,s)$ probability contours.
We plot the resulting constraints upon the power spectrum growth in
Figure 2. Here, we have used the probabilities including the galaxies
without known redshifts, as we found above that the constraints from
the redshift sample are weak. Also plotted is the growth prediction
for $\Lambda$CDM $(\Omega_\Lambda=0.7, \Omega_m=0.3)$ with
$\sigma_8=0.7$ (c.f. section 4.3); we see that the $\Delta^2$
constraints are consistent (within 2$\sigma$) with this model, for
$0<z<1$. Agreement is better between the model and an individual
field, rather than the fields combined, as discussed above. Note that
a $k=14$Mpc$^{-1}$ has been chosen for this plot; the
phenomenological model used presents us with an average,
$k$-independent redshift evolution, as the model has $k$ and $z$ as
independent variables.

The growth of the power spectrum shown on Figure 2 is skewed towards
achieving high amplitude at the present day, with corresponding large
$s$, simply because our small dataset cannot currently differentiate
well between slow-growth low-amplitude and rapid-growth high-amplitude
models. Since there is a large area in parameter space representing
the latter, the overall constraints favour these while not excluding
slower, $\Lambda$CDM growth. Larger surveys will be required in order
to improve the measurement of matter power spectrum growth, by
measuring the growth of the shear correlation function more accurately
as a function of redshift and thus restricting the $A-s$
degeneracy. With these larger surveys, the procedure we have outlined
will allow detailed study of the collapse of structures in the
Universe.

In this section, we have measured the growth rate and amplitude of the
matter power spectrum using a phenomenological model. However, this is
only one approach to the extraction of useful information from a 3-D
shear field; instead of fitting a phenomenological model as above, we
can directly fit the shear predictions for a range of cosmological
parameters. We will carry out this analysis in the following section.

\begin{figure}
\psfig{figure=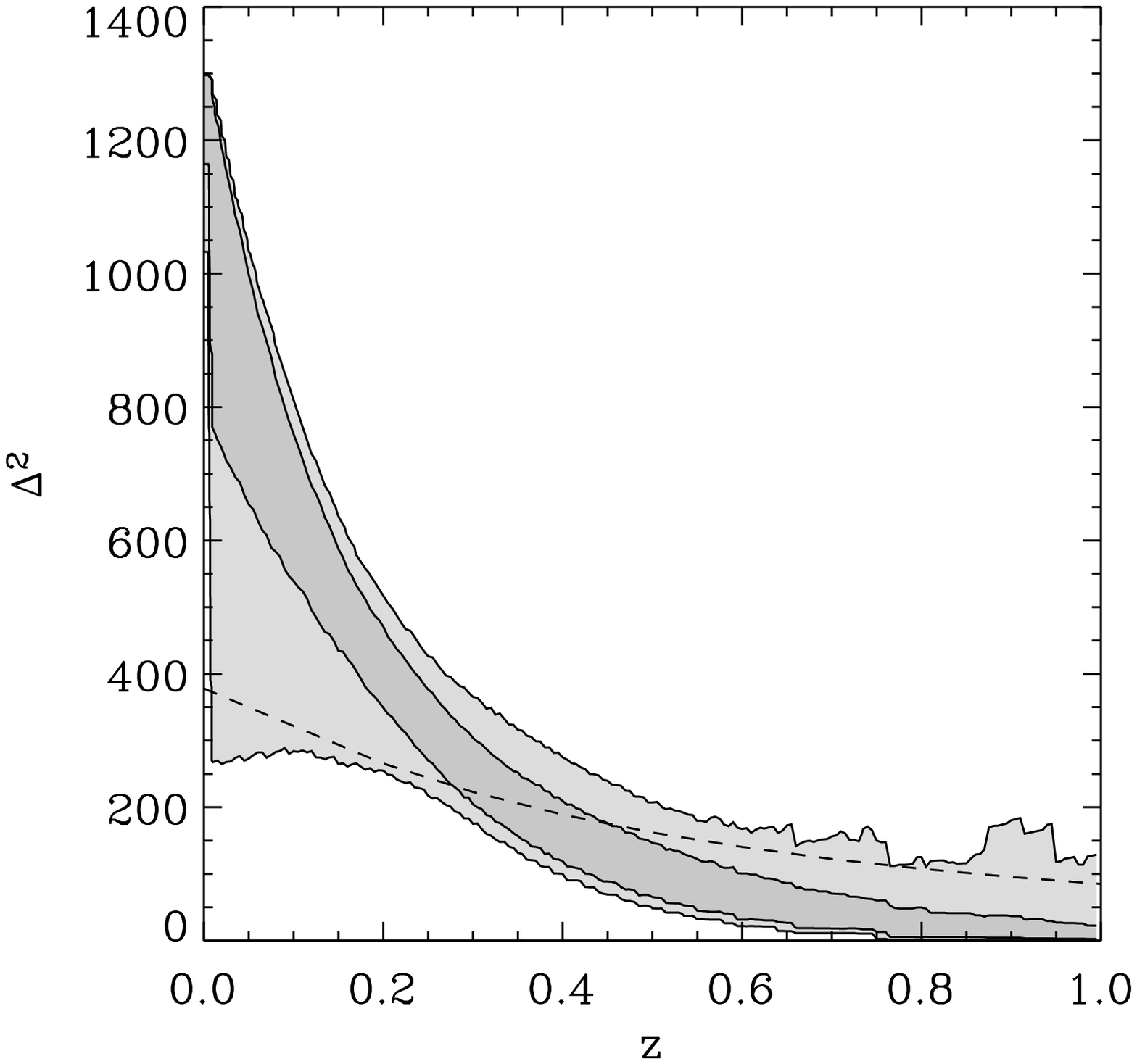,height=7cm,angle=0}
\psfig{figure=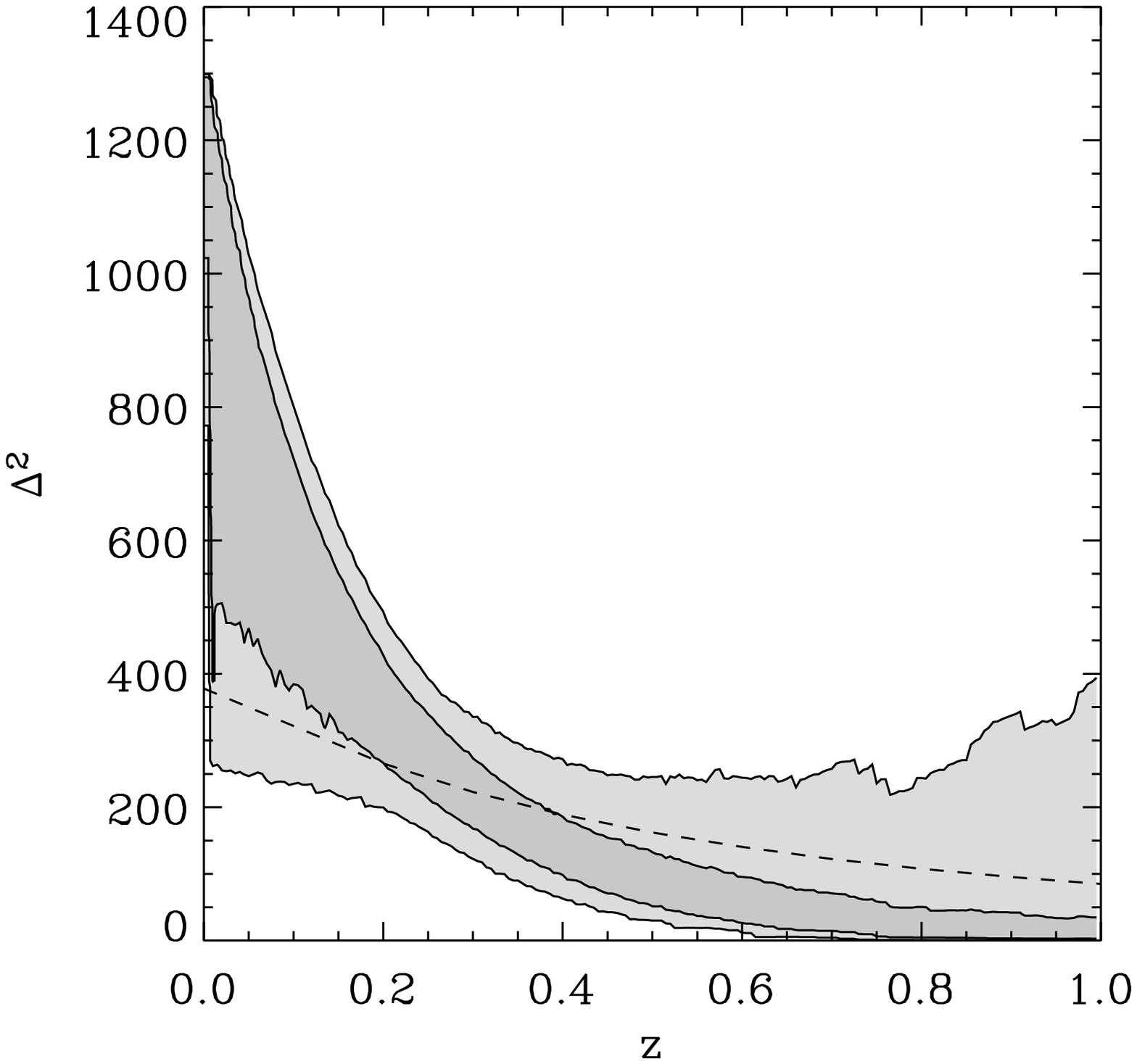,height=7cm,angle=0}
\psfig{figure=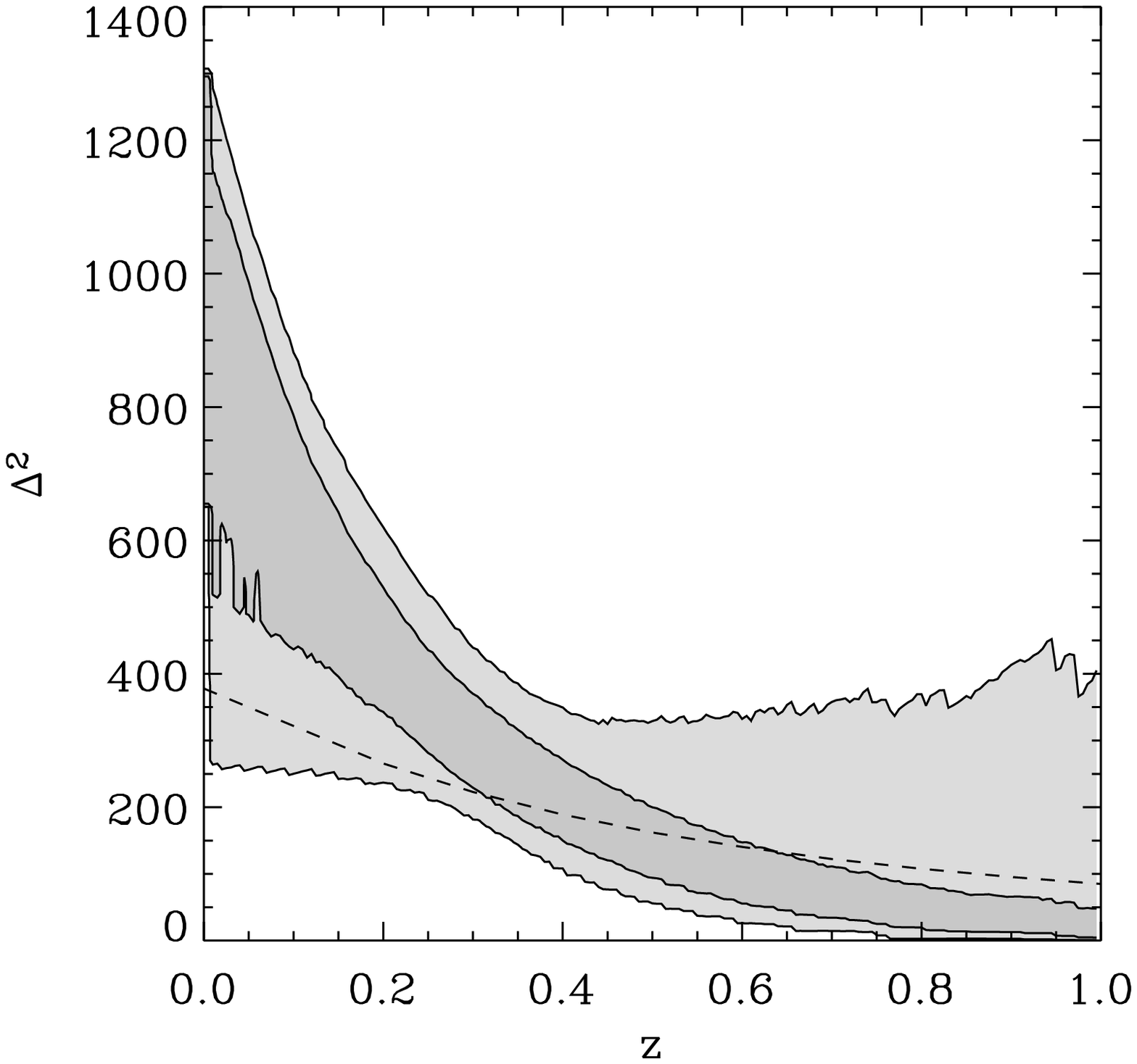,height=7cm,angle=0}
\caption{Constraints on power spectrum as a function of redshift $z$,
displayed for $k=14$Mpc$^{-1}$. The contours show 1 and
2$\sigma$ confidence for the behaviour of the power spectrum at a
given redshift. Also plotted (dashed line) is the prediction for
$\Lambda$CDM with $\sigma_8=0.7$. Top panel: results for CDFS and S11
combined; middle panel: results for CDFS only; bottom panel: results
for S11 only.}
\end{figure}

\subsection{Constraints on the cosmological constant}

Instead of a direct parameterisation of the slope of the power
spectrum with redshift, we can characterise the 3-D shear data by the
constraints which they provide on cosmological parameters. With the
current small dataset, it is not possible to obtain cosmologically
interesting constraints on many parameters at once. Instead, here we
will use priors from WMAP for most parameters, and examine the
resulting constraints on the normalisation $\sigma_8$ and cosmological
constant $\Lambda$ from the COMBO-17 data. Future work with large 3-D
shear datasets will be able to obtain estimations of full parameter
sets from shear alone.

\subsubsection{Non-flat Models}

We choose in the first instance to use the Hubble Key Project and WMAP
measurements of parameters $H_0=72\pm5$km s$^{-1}$ Mpc$^{-1}$ and
$\Omega_b=0.047\pm0.006$ (Freedman et al 2001, Spergel et al 2003),
and examine the slice in parameter space at $\Omega_m=0.3$ allowing
$\Omega_\Lambda$ and $\sigma_8$ to vary (i.e. we will allow non-flat
models, $\Omega_\Lambda+\Omega_m \ne 1$). This will enable us to
examine the constraint from our 3-D lensing method on $\Omega_\Lambda$
if $\Omega_m$ is known. We calculate the 3-D shear correlation
functions for this set of parameters as described in Section 2. We
then make $\chi^2$ fits to the data as above, varying $\sigma_8$ in
steps of 0.05 and $\Omega_\Lambda$ in steps of 0.1.

The resulting weak constraints on these parameters, while either
excluding or including the unknown redshift sample fixed at $z=0.95$
(with $\pm0.05$ marginalisation to account for the uncertainty in
median redshift, plus marginalisation over $H_0$), are shown in Figure
3. In the case where we use only galaxies with known redshift, we
obtain only very broad constraints on $\sigma_8$ and
$\Omega_\Lambda$. We find an upper limit on $\Omega_\Lambda$ of
$\Omega_\Lambda<4.0-4.2 \sigma_8$ at the $1\sigma$ level, or
$\Omega_\Lambda<5.15-4.8\sigma_8$ at the $2\sigma$ level. These
results also constrain $\sigma_8$ to have a low normalisation,
$\sigma_8<0.70 (0.93)$ at the $1 (2) \sigma$ confidence level for
$\Omega_\Lambda=0.7$. This is consistent with the $\sigma_8$
normalisation found for the COMBO-17 dataset for a 2D cosmic shear
analysis (Brown et al 2003). Perhaps this is not surprising for the
two fields of interest here, as both are rather empty fields
containing little evidence of substantial large-scale structure, other
than the cluster in one chip of the S11 field. Therefore further
fields will require analysis before one can obtain a definitive value
for $\sigma_8$ from this method.

If we include galaxies with unknown redshifts, assigned a redshift
$0.95\pm0.05$, our constraints remain weak but show that the
$\Omega_\Lambda$ estimate does vary weakly with $\sigma_8$; the best
fit joint constraint line is
$\Omega_\Lambda=5.6-7.7(\sigma_8\pm0.07)$. This behaviour is as
expected; the weakness of the dependence shows that $\Lambda$CDM and
OCDM models have quite similar evolution over $0<z<1$. As seen above, these
results give a low estimate of $\sigma_8$ on our fields; note that we
already expected these fields to give a low estimate for the spatial
fluctuation of dark matter.

\begin{figure}
\psfig{figure=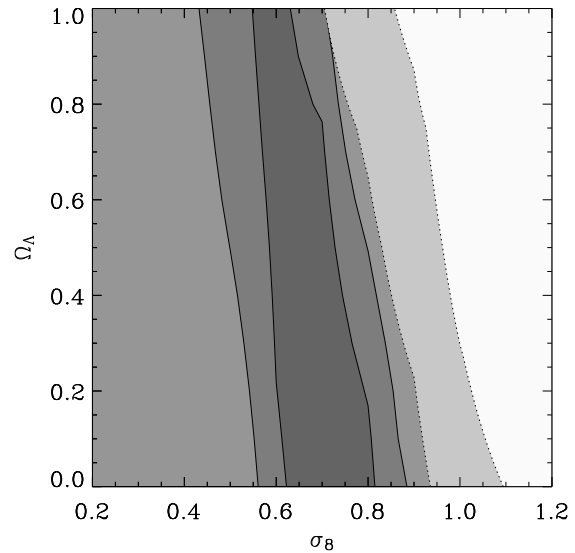,width=\columnwidth,angle=0}
\caption{Constraints on $\Omega_\Lambda$ and $\sigma_8$ from the
COMBO-17 survey, with $\Omega_m$ fixed to be 0.3 and curvature allowed
to vary arbitrarily. Dotted lines (and lightest two greyscales):
constraints solely from sample with known redshifts. Solid lines (and
darker two greyscales): constraints including galaxies with unknown
redshift.}
\end{figure}

\subsubsection{Flat Models}

An alternative means of deriving useful constraints from this small
dataset is to fix $\Omega_\Lambda + \Omega_m$=1, i.e. we can impose
the flatness condition measured by WMAP ($\Omega_{\rm tot}=1.02\pm.02$
with Hubble key project $H_0$ prior, Freedman et al 2001, Spergel et
al 2003). With this condition, we can again make shear correlation
function predictions, then obtain $\chi^2$ fits to the data for
various values of $\sigma_8$ and $\Omega_\Lambda$, incrementing in
units of 0.05 and 0.05 respectively. We marginalise over $H_0$ and
$z_{\rm median}$ as before.

The results are shown in Figure 4, for the cases where we exclude and
include the unknown redshift sample fixed at $z=0.95\pm0.05$. Here we
find that the constraints on $\Omega_\Lambda$ or $\Omega_m$ are
substantially improved over the case where we do not prescribe
flatness. 

In the case where we exclude the unknown redshift galaxies, we find a
weak constraint $\Omega_\Lambda>1-0.16\sigma_8^{-1.95} (1\sigma)$ or
$\Omega_\Lambda>1-0.25\sigma_8^{-1.76} (2\sigma)$. If we then force a
prior $\sigma_8=0.84\pm0.04$, we obtain
$\Omega_\Lambda=0.91^{+.09}_{-0.27}$ at the $2\sigma$ level,
consistent with standard $\Lambda$CDM as defined above.

Again, these results offer a low normalisation of the matter power
spectrum; for $\Omega_\Lambda=0.7$, we require $\sigma_8<0.72$ at the
$1\sigma$ level. As discussed in the previous section, this result is
dominated by the fact that our two fields are devoid of
significant large-scale structure; further fields will require
measurement before cosmological implications can be conclusively drawn.

If we now include galaxies with unknown redshift, we find a best-fit
constraint $\Omega_\Lambda=1-0.15(\sigma_8\pm0.04)^{-1.5},
\sigma_8<0.72 (1\sigma)$. In the case where we impose a prior
$\sigma_8=0.84\pm0.04$, we obtain $\Omega_\Lambda=0.83^{+0.06}_{-0.11}
(2\sigma)$. This, and the constraint directly from our contours
$\Omega_\Lambda<0.76 (1\sigma)$ is in accord with the concordance
$\Lambda$CDM model; only the power spectrum normalisation is outside
the range expected.

Despite the limitations due to the size of our dataset, these
constraints are already impressive, and demonstrate the power of using
3-D information. Figure 4 compares the accuracies of the 2D analysis of
Brown et al (2003, dashed lines) for only the CDFS and S11 fields with the
current 3-D analysis (solid lines); clearly one gains very
significantly from the inclusion of the redshift information, by a
factor of $>2$ everywhere in uncertainty (beyond $\sigma_8>0.4$). This
holds out the promise of very precise measurements of cosmological
parameters with future 3-D lensing surveys.

\begin{figure}
\psfig{figure=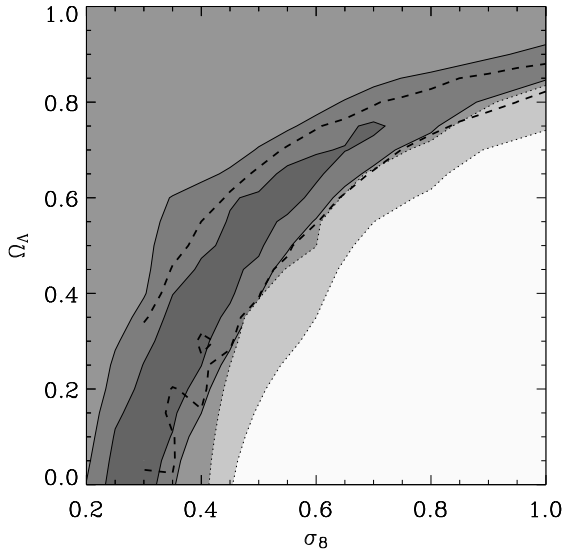,width=\columnwidth,angle=0}
\caption{Constraints on $\Omega_\Lambda$ and $\sigma_8$, with
$\Omega_\Lambda+\Omega_m$ fixed to be 1. Dotted line (and lighter two
greyscales): constraints solely from sample with known
redshifts. Solid line (and darker two greyscales): constraints
including faint sample. Dashed lines show the constraints from our 2D
lensing analysis in Brown et al (2003).}
\end{figure}

\section{Conclusions}

In this paper, we have explored the evolution of large-scale structure
for redshifts $z<1$. This has been achieved using weak gravitational
lensing together with photometric redshifts for a set of galaxies,
both measured upon two fields from the COMBO-17 survey (Wolf et al
2001).

We have described the construction of theoretical models of evolving
matter power spectra, both in terms of phenomenology of the growth of
structure, and in terms of characterisation of this growth with
cosmological parameters. We have then shown how to calculate the shear
power spectrum from this matter power spectrum, and vice versa. In
particular, we have discussed the importance of the cross power
spectrum for shears between redshift shells, and have related
phenomenological models for evolution between the matter and shear
cross power spectra.

In order to practically implement this theoretical framework, we have
described the usefulness of the COMBO-17 survey for 3-D shear
analysis, due to its accurate photometric redshifts for galaxies in
the survey (with $\Delta z\simeq 0.05$ for $0<z<1.0$) and its
well-studied shear catalogue.

We proceeded to apply a least squares estimator analysis to the data,
finding that best-fit matter power spectra included significant
evolution. In particular, we found that the growth rate expected for
$\Lambda$CDM was approximately 50 times more likely than a zero
evolution model, in the case where we ignore galaxies with unknown
redshifts. However, we found that the best-fit rate of growth
understandably depended sensitively upon the present-day amplitude of
the power spectrum. We calculated constraints upon $\Delta^2(z)$,
finding a reasonable match with that predicted by $\Lambda$CDM for
$0<z<1$, but allowing larger present-day amplitude of the power
spectrum with more dramatic growth rates.

As an alternative route to extracting information from the 3-D shear
field, we obtained least squares estimates of the cosmological
constant and the normalisation of the power spectrum. We find low
values of $\sigma_8$ in this small area of sky, with best-fit
constraints $\Omega_\Lambda=5.6-7.7(\sigma_8\pm0.07)$ and
$\Omega_\Lambda=1-0.15(\sigma_8\pm0.04)^{-1.5}, \sigma_8<0.72$ if we
use WMAP priors on $\Omega_m$ or $\Omega_k$ respectively. We find that
the latter constraint is a factor of 2 improvement upon the constraint
obtained by 2D lensing for the same area of sky.

These results demonstrate that 3-D statistical lensing is a useful
means of examining dark matter evolution. If one is interested in
measuring only cosmological parameters, this method is a complementary
alternative to measuring 2-D higher-order statistics of the shear
field; the question of the comparative (and combined) merits of the
two approaches is the subject of ongoing research (c.f. Benabed \& van
Waerbeke 2003, Refregier et al 2003). However, if one seeks a direct
measurement of the dark matter evolution itself, 3-D methods such as
the one explored here are essential.  With large multi-colour imaging
surveys in the next few years such as that of SNAP (Rhodes et al
2004), the potential of this approach will be fully realised,
affording measurements of the cosmological constant and dark energy
equation of state parameters to unprecedented accuracy, together with
precise reconstructions of the growth of the power spectrum as a
function of redshift.

\section*{Acknowledgments}

DJB, MG and MLB are supported by a PPARC Postdoctoral Fellowship; ANT
is supported by a PPARC Advanced Fellowship. CW was supported by the
PPARC Rolling Grant in Observational Cosmology at the University of
Oxford. We thank Alan Heavens and Hans-Walter Rix for extremely useful
discussions.

\label{lastpage}

\end{document}